\documentclass[10pt,a4paper,twoside]{article}
\usepackage{amsfonts,psfig,graphicx}
\usepackage{etc9macro}

\newcommand{\pr}{\partial}
\newcommand{\ub}{\overline{u}}
\newcommand{\pb}{\overline{p}}
\newcommand{\uub}{{\overline{{\bf{u}}}}}
\newcommand{\uu}{{\bf{u}}}
\newcommand{\be}{\begin{equation}}
\newcommand{\en}{\end{equation}}

\begin{document}

\begin{etcpaper}

\begin{etctitle}
  \etcsettitle{Leray simulation of turbulent shear layers}
              {Leray simulation of turbulent shear layers}
  \etcaddauthor{1}{Geurts}{B.J.}
  \etcaddauthor{2}{Holm}{D.D.}
  \etcaddaffiliation{1}{Faculty of Mathematical Sciences, University of Twente,\break
                        P.O. Box 217, 7500 AE Enschede, The Netherlands}
  \etcaddaffiliation{2}{Theoretical Division and Center for Nonlinear Studies, \break
                        Los Alamos National Laboratory, MS B284 Los Alamos, NM 87545, USA}
  \etcemailaddress{b.j.geurts@math.utwente.nl}
  \etcsetpaperid{100}
\end{etctitle}

\section{Introduction}

Accurate modeling and simulation of turbulent flow is a topic of intense ongoing research. Two main approaches are identified, differing
by the amount of detail that is included in the physical and numerical description. Direct numerical simulation (DNS) aims to calculate
the full, unsteady solution to the governing Navier-Stokes equations. While accurate in principle, DNS is severely restricted by limitations
in spatial resolution. This situation summons alternative simulation approaches that are aimed at capturing the primary features of the
flow above a certain length-scale only. A prominent example of this is the large-eddy simulation (LES) strategy in which a smoothing of the
flow features and a corresponding reduction in the flow complexity is introduced by spatial filtering, at the expense of introducing a
`subgrid' closure problem.

We consider so-called Leray regularization of the convective contributions \cite{leray}. This gives rise to a subgrid parameterization
which involves both explicit filtering and (approximate) inversion. The Leray model also arises from the $\alpha$-modeling strategy
derived via Kelvin's circulation theorem \cite{geurtsholm2002a}. We study the dynamics associated with the Leray model in a turbulent
mixing layer and compare predictions with filtered DNS results and findings due to dynamic (mixed) models \cite{vreman_jfm}. In particular,
the kinetic energy, momentum thickness and energy-spectra are analyzed, establishing favorable performance of the Leray model and
robustness at arbitrarily high Reynolds number. This is unique for a similarity-type model that does not contain an explicit eddy-viscosity
term.

We provide the basic Leray formulation in section~\ref{basic} together with the numerical inversion of the filter. Application to
turbulent mixing is presented in section~\ref{turbshear} and concluding remarks are collected in section~\ref{concl}.

\section{Basic Leray formulation}
\label{basic}

In the filtering approach the evolution of the filtered solution $\{ \ub_{i},~\pb \}$ is governed by the spatially smoothed Navier-Stokes
equations. Filtering the nonlinear terms gives rise to the turbulent stress tensor $ \tau_{ij}={\overline{u_{i}u_{j}}} -\ub_{i} \ub_{j}$.
Here $\ub_{i}=L(u_{i})$ is the filtered velocity field in the $x_{i}$ direction, with $L$ denoting the linear filtering operation which we
assume to have a (formal or approximate) inversion $L^{-1}$. Likewise, $\pb=L(p)$ is the filtered pressure. Expressing $\tau_{ij}$ in terms
of the filtered velocity is the basic closure problem in LES. 

Leray regularization provides an intuitively appealing method for modeling $\tau_{ij}$. In this formulation the convective fluxes are
replaced by $\ub_{j}\pr_{j}u_{i}$, i.e., the solution $\uu$ is convected with a smoothed velocity $\uub$. Consequently, the nonlinear
effects are reduced by an amount governed by the smoothing properties of $L$. For commuting filters $L$ the governing equations in the
Leray formulation can be written as
\begin{equation}
\pr_{j}u_{j}=\pr_{j}\ub_{j}=0~~;~~\pr_{t}u_{i}+\ub_{j}\pr_{j}u_{i}+\pr_{i}p-\frac{1}{Re}\pr_{jj}u_{i}=0
\label{leray1}
\end{equation}
Uniqueness and regularity of the solution to these equations have been established rigorously \cite{leray}. The Leray formulation contains
the unfiltered Navier-Stokes equations as a limiting case. The unfiltered solution can readily be eliminated, giving rise to a closed
formulation for $\{ \ub_{i},\pb \}$:
\begin{equation}
\pr_{t}\ub_{i} + \pr_{j}(\ub_{j}\ub_{i}) + \pr_{i} \pb - \frac{1}{Re} \pr_{jj} \ub_{i} + \pr_{j}m_{ij}^{L}=0
\label{leray2}
\end{equation}
where the asymmetric, filtered similarity-type Leray model $m_{ij}^{L}$ arises:
\begin{equation}
m^{L}_{ij}=L\Big(\ub_{j}L^{-1}(\ub_{i})\Big) -\ub_{j}\ub_{i}
\label{basic_leray}
\end{equation}

In LES, one commonly adopts compact support filters with filter-kernel $G$. In one spatial dimension such filtering can be expressed as
\begin{equation}
\ub(x)=L(u)=\int_{-\Delta/2}^{\Delta/2} G(z) u(x+z)~dz
\end{equation}
where $\Delta$ is the filter-width. In actual simulations the resolved fields are known only at a set of grid
points $\{ x_{m} \}_{m=0}^{N}$. Numerical filtering corresponds to kernels
\begin{equation}
G(z)=\sum a_{j} \delta(z-z_{j})~~;~~|z_{j}| \leq \Delta/2
\end{equation}
We consider three-point filters with $a_{0}=1-\alpha$, $a_{1}=a_{-1}=\alpha/2$ and $z_{0}=0$, $z_{1}=-z_{-1}=\Delta/2$. In addition,
we use $\alpha=1/3$ which corresponds to Simpson quadrature of the top-hat filter. The application of $L^{-1}$ to a general discrete
solution $\{ u(x_{m})\}$ can be specified using discrete Fourier transformation as \cite{kgvg99}
\be
L^{-1}(u_{m})={{\sum_{j=-n}^{n}}} \Big( \frac{\alpha-1+\sqrt{1-2\alpha}}{\alpha} \Big)^{|j|} (1-2\alpha)^{-1/2} {{u_{m+rj/2}}}
\en
where the subgrid resolution $r=\Delta/h$ (with $h$ the grid-spacing) is assumed to be even. An accurate and efficient inversion can be
obtained with only a few terms, recovering the original signal to within machine accuracy. At fixed $\Delta$, variation of the subgrid
resolution $r$ allows an independent control over flow-smoothing and numerical representation \cite{geurtsfroehlich2002}.

\section{Shear layers at arbitrary Reynolds number}
\label{turbshear}

The turbulent mixing layer \cite{vreman_jfm} is simulated in a volume $L^{3}$ at various $Re$. At a modest $Re=50$, an
assessment of the quality of Leray modeling is obtained by comparing with filtered DNS data and dynamic models.
Moreover, we consider this flow at high $Re$, adopting a fourth order accurate spatial discretization.

\begin{figure}[htb]

\centerline{
\psfig{figure=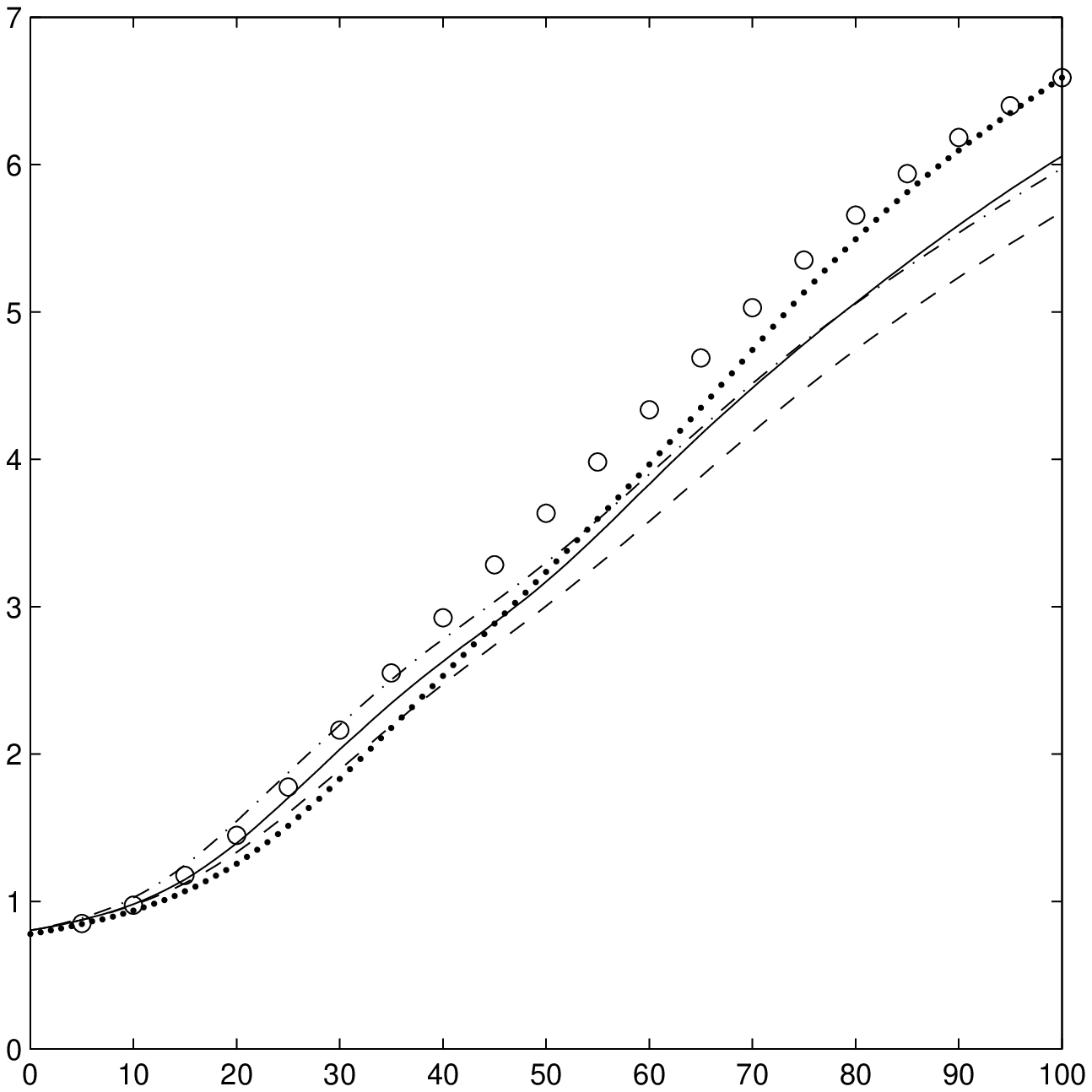,width=0.4\textwidth} (a)
\psfig{figure=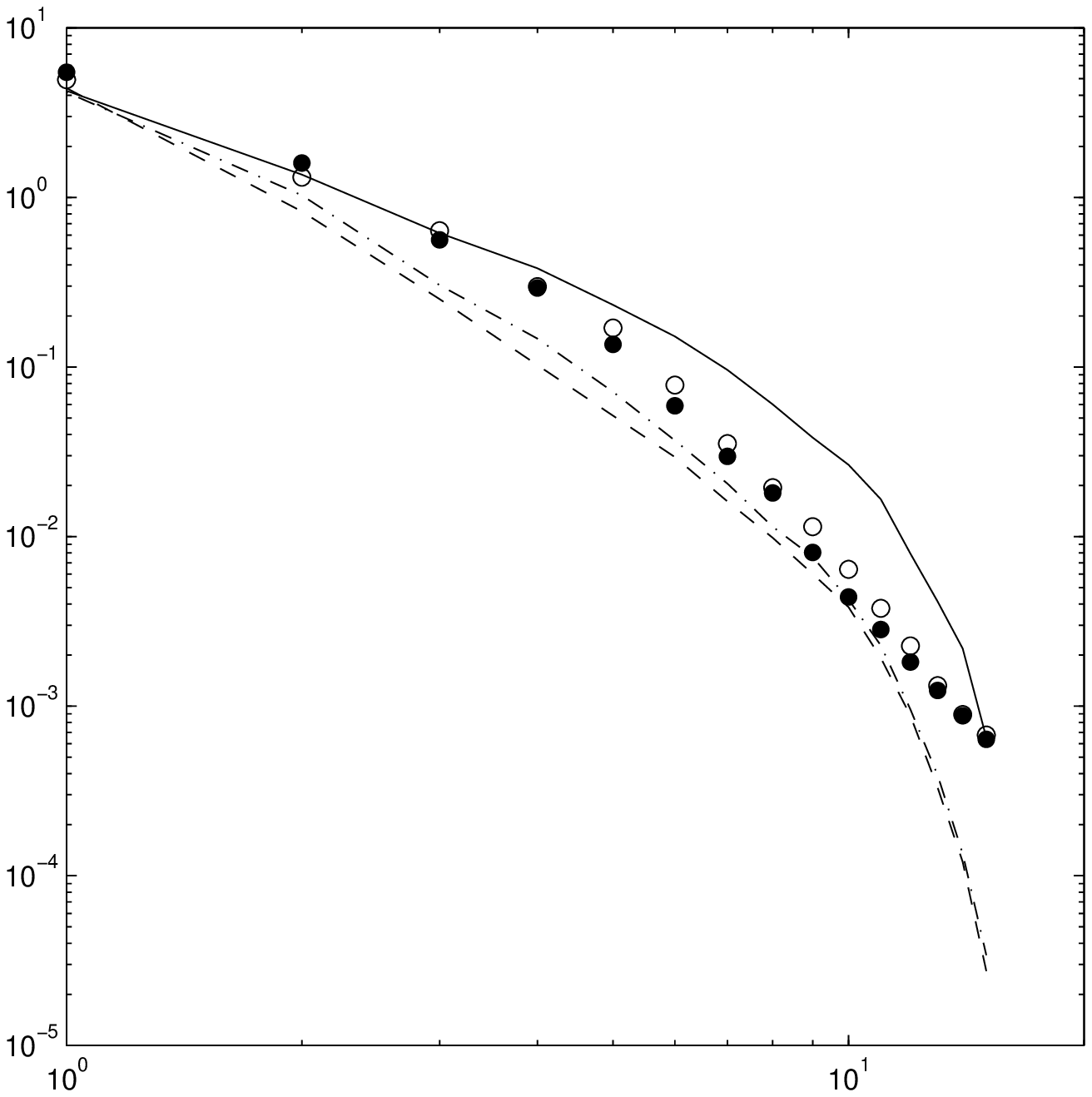,width=0.4\textwidth} (b)
}

\vspace*{-0.35\textwidth}

\hspace*{0\textwidth} $\theta$ \hspace*{0.425\textwidth} $A$

\vspace*{0.3\textwidth}

\hspace*{0.35\textwidth} $t$ \hspace*{0.4\textwidth} $k$

\vspace*{-3mm}

\caption{
Momentum thickness $\theta$ (a) and streamwise kinetic energy spectrum $A$ at $t=75$ (b): Leray ($32^3$: solid, $64^3$: dotted), dynamic
model ($32^3$: dashed), dynamic mixed model ($32^3$: dash-dotted), filtered DNS ($\circ$). LES at $\Delta=L/16$.}

\label{fig1}
\end{figure}

Visualization of the DNS, obtained on $192^{3}$ cells, demonstrates a well developed flow beyond $t=40$. In Fig.~\ref{fig1}(a)
the evolution of the momentum thickness shows Leray predictions to compare well with filtered DNS data and with dynamic (mixed)
models at $32^{3}$. This is improved using $64^{3}$ points. In Fig.~\ref{fig1}(b) the streamwise
kinetic energy spectrum at $t=75$ on $32^3$ shows improved capturing of the large and intermediate scales, compared to dynamic models. The
dissipation of the smaller scales is significantly improved on $64^3$. The Leray model shows both forward and backward
scattering of energy \cite{geurtsholm2002a}.

\begin{figure}[htb]

\centerline{
\psfig{figure=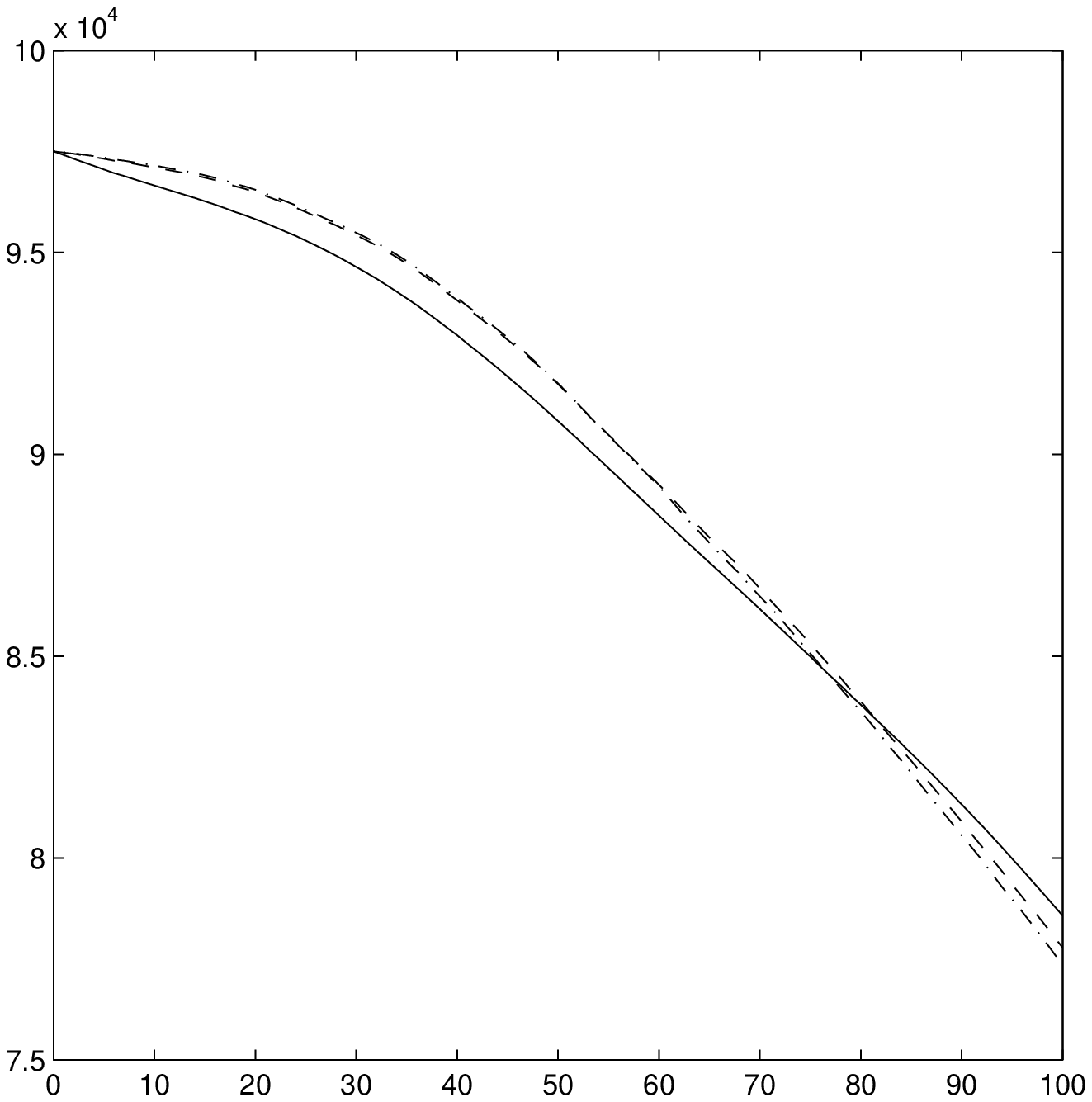,width=0.4\textwidth} (a)
\psfig{figure=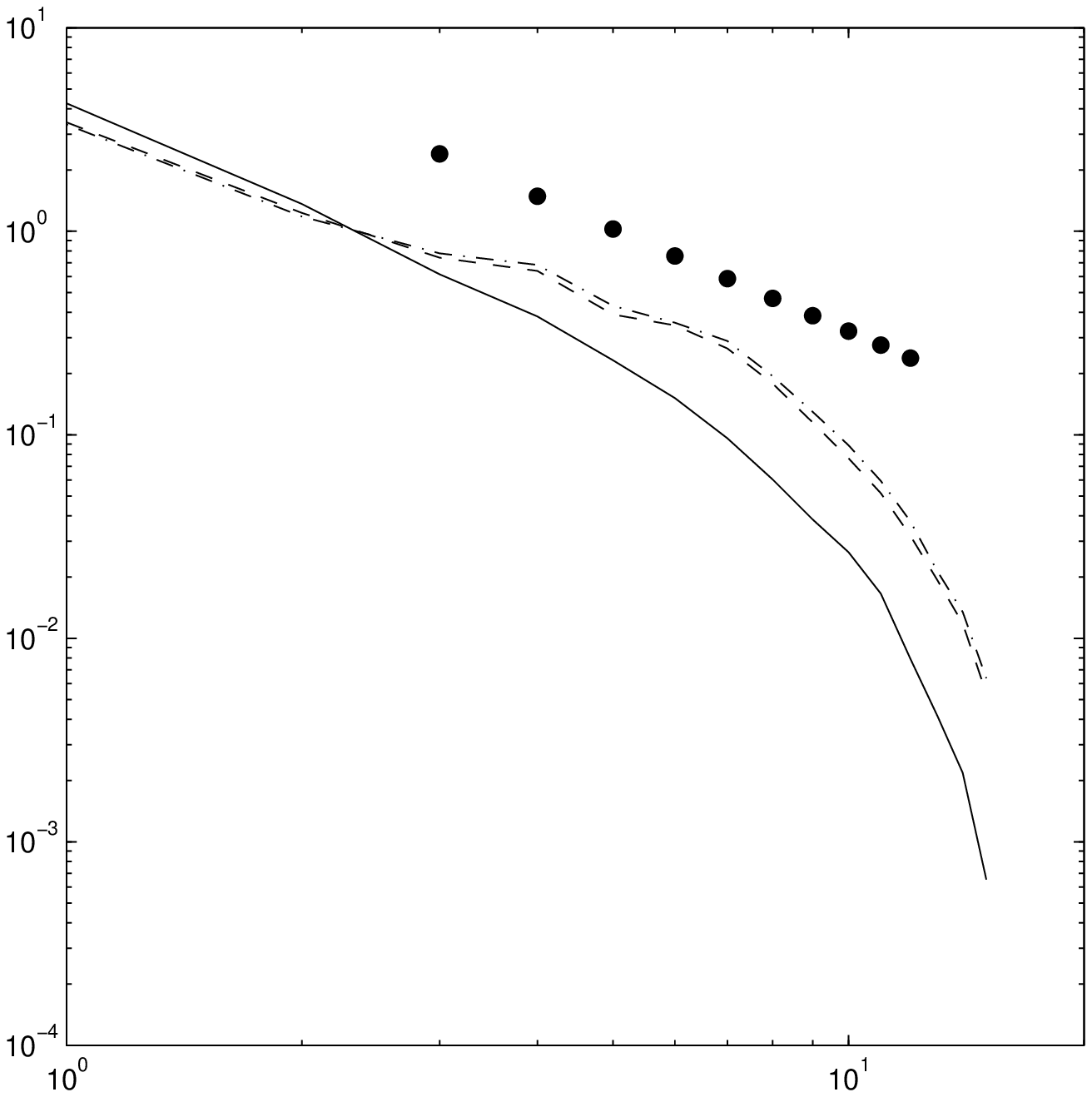,width=0.4\textwidth} (b)
}

\vspace*{-0.35\textwidth}

\hspace*{0\textwidth} $E$ \hspace*{0.425\textwidth} $A$

\vspace*{0.3\textwidth}

\hspace*{0.35\textwidth} $t$ \hspace*{0.4\textwidth} $k$

\vspace*{-3mm}

\caption{
Kinetic energy $E$ (a) and streamwise kinetic energy spectrum $A$ at $t=75$ (b): Leray model at $Re=50$ (solid), $Re=500$ (dashed),
$Re=5000$ (dash-dotted). LES at $32^{3}$ and $\Delta=L/16$. The dotted line represents $k^{-5/3}$.}

\label{fig2}
\end{figure}

A particularly appealing property of Leray modeling is the robustness at very high and even infinite Reynolds number, cf. Fig.~\ref{fig2}
(a,b). Although comparison with filtered DNS data is impossible, we observed that the smoothed flow dynamics is properly captured and a
nearly grid-independent solution is obtained as $r=\Delta/h \geq 4$ \cite{geurtsfroehlich2002}. The kinetic energy is likely to be
slightly over-predicted. At high $Re$ the spectrum displays a region with $k^{-5/3}$ behavior.

\section{Concluding remarks}
\label{concl}

The Leray model was found to predict the momentum thickness properly while exhibiting both forward and backward transfer of energy. Further
analysis shows reliable levels of turbulence intensities and correct behavior of kinetic energy. The Leray model has a tendency to
underestimate dissipation. The computational overhead associated with the Leray model is lower than that of dynamic (mixed) models and no
introduction of {\it ad hoc} parameters is required. The regularized dynamics shows an appealing robustness at high $Re$.

\end{etcpaper}

\begin{thebibliography}{99}
\bibitem{leray}
J.~Leray.
\newblock Sur les movements d'un fluide visqueux remplaissant l'espace.
\newblock {\em Acta Mathematica}, 63: 193--248, 1934.

\bibitem{geurtsholm2002a} 
B.J. Geurts, D.D. Holm.
\newblock Alpha-modeling strategy for LES of turbulent mixing.
\newblock {\it Turbulent flow computation}, Eds: D. Drikakis, B.J. Geurts. Kluwer Academic Publishers. To appear. 2002.

\bibitem{vreman_jfm} 
A.W. Vreman, B.J. Geurts, J.G.M. Kuerten.
\newblock  Large-eddy simulation of the turbulent mixing layer.
\newblock {\it J. Fluid Mech.} {339}: 357, 1997.

\bibitem{kgvg99} 
J.G.M. Kuerten, B.J. Geurts, A.W. Vreman, M. Germano.
\newblock Dynamic inverse modeling and its testing in large-eddy simulations of the mixing layer.
\newblock {\it Phys. Fluids}, 11: 3778-3785, 1999.

\bibitem{geurtsfroehlich2002}
B.J. Geurts, J. Fr\"ohlich.
\newblock Numerical effects contaminating LES; a mixed story.
\newblock {\em Modern strategies for turbulent flow simulation}, Ed: B.J. Geurts. Edwards Publishing. 317-347, 2001.

\end{thebibliography}
\end{document}